\documentclass{nature}
\usepackage{hyperref}
\usepackage{graphicx}
\usepackage{multibib}
\usepackage{amssymb,amsmath}
\usepackage{color}

\title{The paradoxes of the Late Hesperian Mars ocean}

\author{M. Turbet$^{1}$ \& F. Forget$^1$}

\makeatletter
\let\saved@includegraphics\includegraphics
\AtBeginDocument{\let\includegraphics\saved@includegraphics}
\renewenvironment*{figure}{\@float{figure}}{\end@float}
\makeatother

\begin{document}

\maketitle

\begin{affiliations}
 \item Laboratoire de M\'et\'eorologie Dynamique, IPSL, Sorbonne Universit\'es, UPMC Univ Paris 06, CNRS, 4 place Jussieu, 75005 Paris, France.
\end{affiliations}

\begin{abstract}

The long-standing debate on the existence of ancient oceans on 
Mars has been recently revived by 
evidence for tsunami resurfacing events that date from the Late Hesperian 
geological era. It has been argued that these tsunami events 
originated from the impact of large meteorites on a deglaciated or nearly deglaciated ocean 
present in the northern hemisphere of Mars. Here we show that 
the presence of such a late ocean faces a paradox.
If cold, the ocean should have been entirely frozen shortly after its formation, thus preventing the formation of tsunami events.
If warm, the ice-free ocean should have produced fluvial erosion of Hesperian Mars terrains 
much more extensively than previously reported.
To solve this apparent paradox, we suggest a list of possible tests and scenarios that could help to reconcile 
constraints from climate models with tsunami hypothesis. These scenarios could be tested in future dedicated studies.

\end{abstract}

\section{Introduction}

The existence of liquid water oceans on ancient Mars has long been a topic of 
debate\cite{Parker:1989,Baker:1991,Malin:1999,Carr:2003,Perron:2007,Head:2018} and has strong 
implications for the search for life in the solar system. 
Specifically, the lack of wave-cut paleoshore-line features and the presence 
of lobate margins seem to be inconsistent with the presence of a Late Hesperian ocean 
(see \cite{Rodriguez:2016} and references therein). 
A review of the historical Late Hesperian ocean controversy, as well as alternative scenarios 
to explain geologic observations are proposed in \cite{Head:2018}.
Recently, two studies\cite{Rodriguez:2016,Costard:2017} independently identified the presence of 
highland-facing lobate debris deposits in Arabia Terra, along the dichotomy boundary, 
interpreted as tsunami deposits.
The overlap of several distinct lobate deposits as well as their wide range of elevation suggest 
the possibility of multiple (at least two) tsunami events\cite{Rodriguez:2016,Costard:2017}. Both studies\cite{Rodriguez:2016,Costard:2017} 
reported that these tsunami events were likely 
caused by the collision of large meteorites (3-6~km in diameter) on an ice-free or sea-ice covered 
ocean located in the northern lowlands of 
Mars.

Therefore, Mars could have hosted a large body of liquid 
water hundreds of millions of years later than the formation of the valley networks. Below 
we discuss the implications of the presence of a deglaciated (or nearly deglaciated) ocean on both the 
atmosphere and the geology of Late Hesperian Mars.



\section{The paradoxes of a cold ocean}

Sustaining a liquid-water ocean, even ice-covered, would require a very strong 
greenhouse effect involving a mixture of greenhouse gases. 
3-D climate modeling of early Mars under an atmosphere composition of 
only CO$_2$ and H$_2$O, 
performed with a water 
cycle that includes water vapor and clouds, is unable to 
maintain significant amount of liquid water anywhere on 
the red planet, even when maximizing the greenhouse effect of H$_2$O and CO$_2$ ice 
clouds\cite{Forget:2013,Wordsworth:2013}. This major result holds independently of the specific CO$_2$ 
atmospheric content, water content and obliquity. More specifically, an initially warm 
northern ocean possibly fed by outflow channel formation events should freeze extremely rapidly 
under a Late Hesperian Mars CO$_2$ atmosphere\cite{Kreslavsky:2002,Turbet:2017icarus}. For instance, a 303~K, 
200m deep ocean (under a 0.2~bar CO$_2$ atmosphere, at 45$^{\circ}$ obliquity) would 
become entirely ice-covered after $\sim$~1~martian year, and frozen solid 
after $\sim$~4$\times$10$^3$~martian years\cite{Turbet:2017icarus}. 
More generally, a 
northern plains ocean would completely freeze within $10^4$ years, whatever the obliquity, 
surface pressure of CO$_2$, and whatever the initial size and temperature of the ocean 
assumed\cite{Turbet:2017icarus}. 
The freezing process is particularly efficient at low obliquity 
(because of the low insolation at the North Pole) and low CO$_2$ atmospheric pressure (because of the small greenhouse effect of CO$_2$). At high obliquity and high CO$_2$ surface pressures, 
the poles are warmer, but still too cold to sustain a liquid water ocean (even with ice cover)\cite{Turbet:2017icarus}. 

\begin{figure}
\centering
\includegraphics[width=\linewidth]{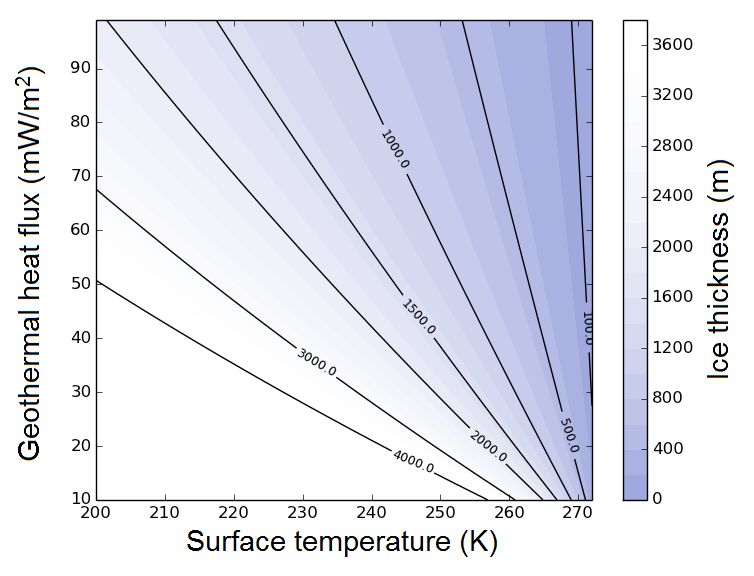}
\caption{Minimum ice thickness covering the ocean calculated as a function of the 
temperature at the top of the sea ice cover and of the geothermal heat flux.}
\label{fig:ice}
\end{figure}

To avoid the need of additional hypothetical greenhouse gases (see below),  
it is tempting to speculate that the ocean could have remained liquid below a
protective ice layer, thanks to a strong geothermal heat flux. 
We calculated the minimum ice thickness (see 
Methods) required to sustain subsurface liquid water, as shown in Fig.~\ref{fig:ice}. Even for  
a geothermal heat flux of 80~mW~m$^{-2}$ (which is an upper limit on 
the geothermal heat flux expected during the Late Hesperian era, and for volcanically active 
regions\cite{Ruiz:2011}), we found that the ice thickness grows very rapidly 
as the surface temperature drops below freezing. The range of conditions that would
have allowed a long-term ice-covered ocean is very narrow: for instance, 
to limit the growth of the ice thickness 
to 100~m or less, the annual mean surface temperature must range 
between 270 and 273~K. 
As a comparison, the minimum ice thickness is $\sim$~1~km for an annual mean surface temperature of 240~K (see Fig.~\ref{fig:ice}) 
which corresponds to the maximum annual mean temperature predicted in the northern lowlands 
by 3-D numerical climate models\cite{Wordsworth:2013,Turbet:2017icarus}.
The presence of salts would have depressed the ocean's water triple point. 
However, even for a drop of 21~K (the lowest freezing point obtainable for a NaCl brine is 252~K\cite{Mohlmann:2011}), 
we calculated that the minimum ice thickness is $\sim$~0.5~km for an annual mean surface temperature of 240~K.


Even with a frozen surface, a Northern ocean on Mars would have been 
unstable on geological timescales. 
If one assumes that a water ocean is formed during a short period of time 
(e.g. by catastrophic outflows) and then left on its own, 
calculations show that it would have eventually 
disappeared after $\sim$~10$^5$~martian years because the water would have 
been progressively transported toward the elevated regions of Mars through sublimation and 
subsequent adiabatic cooling and condensation\cite{Wordsworth:2013,Turbet:2017icarus}. 
Only a thick lag deposit of silicate material\cite{Carr:1990,Mouginot:2012}
 formed on a permanently frozen surface 
could have prevented the water from getting sublimated and migrating 
away to the elevated terrains of Mars. However, even then, it seems almost
unavoidable that the ocean would have frozen down to its bottom, 
as hypothesized in the scenarios where the Vastitas Borealis Formation (VBF) is thought to be 
a remnant of the sublimation residue of a Late Hesperian ocean\cite{Kreslavsky:2002,Carr:2003,Mouginot:2012}.

\section{The paradoxes of a warm ocean} 

Alternatively, we could imagine that for some time the Late Hesperian Martian climate 
was sufficiently warmed by additional strong greenhouse gases,
and thus keeping the ocean at
least partly liquid. For instance, reducing gases (e.g. CH$_4$ and H$_2$) offer a  
way to warm the surface of ancient Mars above the melting point of water\cite{Ramirez:2014,Wordsworth:2017,Kite:2017Nat,Kite:2017collapse,Ramirez:2017}. This effect 
results in part from the strong collision-induced absorptions of CO$_2$-CH$_4$, CO$_2$-H$_2$ and 
H$_2$-H$_2$ pairs\cite{Ramirez:2014,Wordsworth:2017,Turbet:2019}.

However, the persistence of a deglaciated ocean during the Late Hesperian on Mars would raise several
issues. New 3-D Global Climate simulations (see Methods) confirm
that a deglaciated northern ocean could 
be permanently sustained with the assumption of enough greenhouse gases (CO$_2$ and H$_2$ here). 
However, ocean waters would evaporate rapidly and subsequently migrate toward the elevated Martian terrains 
through the mechanism of adiabatic cooling mentioned above for the case of a frozen ocean. 
This process is rapid because evaporation and sublimation rates increase 
exponentially with temperature. 
For instance, a 200m deep deglaciated northern ocean would completely evaporate within $\sim$~10$^3$~martian years, 
whatever the obliquity, surface pressure of CO$_2$ and of additional reducing gases, 
and whatever the initial temperature ($>$~273~K) of the ocean assumed.

\begin{figure}
\centering
\includegraphics[width=\linewidth]{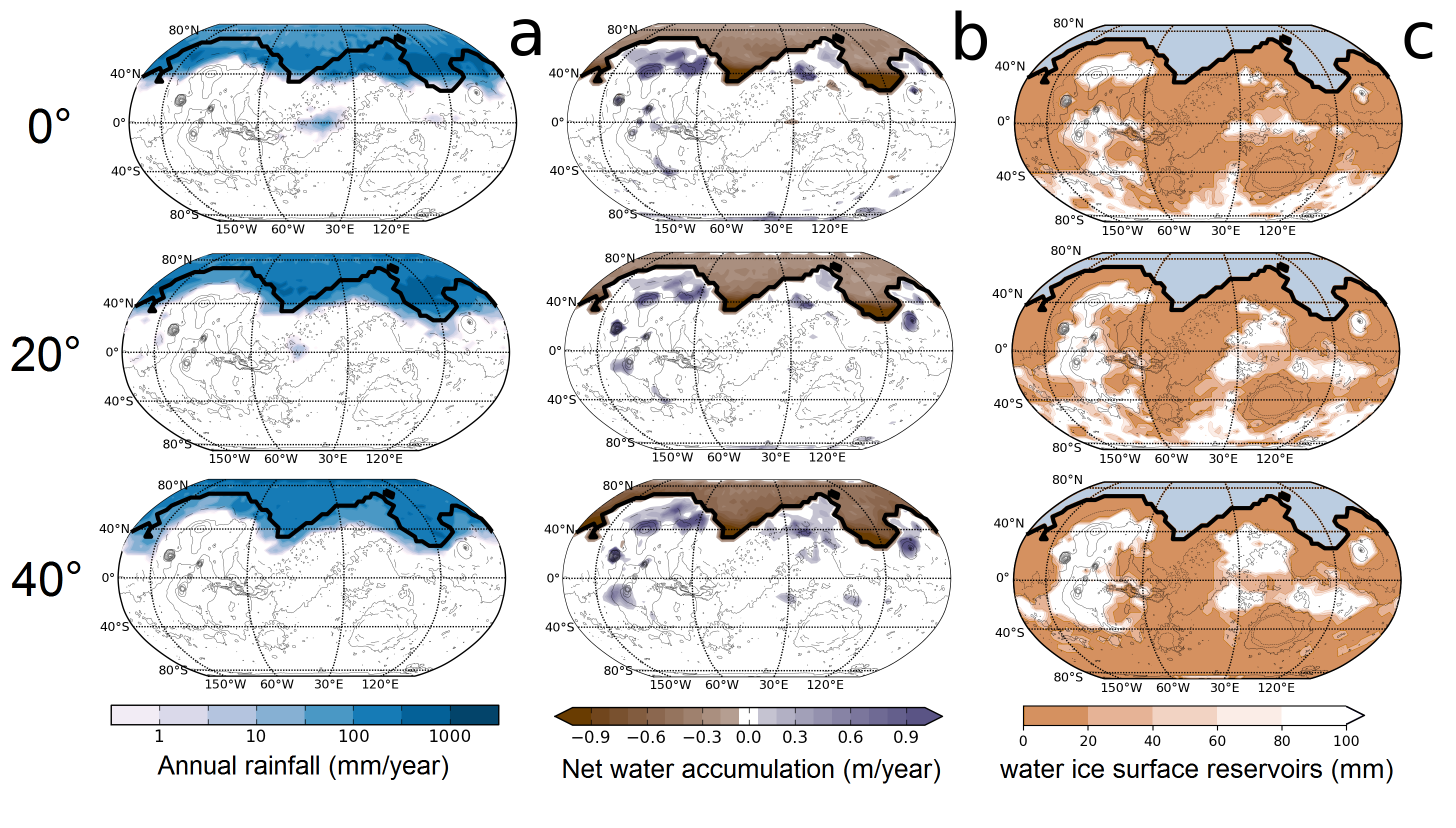}
\caption{Annual cumulated rainfall (left panel), annual net surface accumulation of water 
(middle panel), and position of permanent ice reservoirs (right panel). These figures are based on 
the results of 3-D Global Climate Model (GCM) simulations with a thick CO$_2$ atmosphere and H$_2$, with 
a permanent deglaciated ocean (indicated by a thick black line; the location of the ocean is consistent 
with previous estimate\cite{Costard:2017}), and at three different obliquities (0, 20 and 40$^{\circ}$). We used the present-day Mars 
MOLA topography, indicated by grey contour lines.}
\label{fig:ocean}
\end{figure}

In the simulations, a part of the atmospheric water 
returns to the surface as rain, near the ocean shoreline (see 
Fig.~\ref{fig:ocean}a). 
Such precipitation would produce extensive fluvial erosion, 
in particular in the regions where evidence for tsunami 
events have been reported\cite{Rodriguez:2016,Costard:2017}, and as 
long as the deglaciated northern ocean remains.
The remainder is sequestered as ice on the elevated terrains (see Fig.~\ref{fig:ocean}c). 
In any case, for the northern ocean to survive, an intense hydrological cycle had to occur 
in order to replenish the water that was transported to the elevated terrains.
Although previous regional maps seem to indicate an absence of 
such a strong hydrological cycle in the mid and late Hesperian geological record (see \cite{Fassett:2011} and references therein), 
this prediction could be tested in more details through further high-resolution geological investigations, 
in particular along the proposed paleo-ocean shoreline.

To solve this paradox, one hypothesis
could be for the ocean to be replenished by groundwater. In this scenario, 
water that condensed on the elevated Martian volcanic regions would have formed thick glaciers 
that would undergo melting at their base, possibly introducing the meltwater into 
subsurface aquifers\cite{clifford:2001}. These subsurface liquid water reservoirs could then 
have provided the water that carved the outflow channels, thus replenishing the northern ocean. 
Such an hypothesis would be consistent with our 3-D Global Climate simulations (see Fig.~\ref{fig:ocean}b) in which 
water tends to condense preferentially close to the regions that sourced the outflow
channels. However it is difficult to reconcile this hypothesis with the estimated
lifetime of the ocean. It has been reported that at least two large 
tsunami events were produced by bolide impacts, resulting in craters 30-50~km in diameter\cite{Rodriguez:2016,Costard:2017}. 
Based on the crater frequency rates of Rodriguez\cite{Rodriguez:2016}, the rate of 
Late Hesperian marine impacts producing craters $\sim$30~km in diameter is one every 2.7~million years.
Unless the tsunamis were the result of very unlikely occurrences, 
the ancient ocean would have to survive for a period of at least a few million years 
to produce the reported two consecutive tsunami events\cite{Rodriguez:2016,Costard:2017}. 
This is also supported by the detection of glacier valleys cross cuting 
first, older tsunami deposits and having floors partly covered by younger, second tsunami deposits, indicating that the time 
gap between the two tsunami events was geologically significant\cite{Rodriguez:2016}.
We estimate from our 3-D Global Climate simulations 
(see Fig.~\ref{fig:ocean}b) that the net evaporation rate of the ocean is at least 0.6~m per martian year, 
and that at least 10$^4$~km$^3$ of water would be required for replenishment per martian year in order for the 
ocean to remain stable. Thus, as much as 1.5$\times$10$^{10}$~km$^3$ (e.g. $\sim$~100~km 
GEL) of water would need to have flown through the outflow channels for a deglaciated northern ocean 
to survive for 2.7~million years. This amount is several orders of magnitude larger than previous 
estimates of the total amount of water required to erode all the Martian outflow 
channels\cite{Carr:2015}.

This paradox could be partly overcome if the Late Hesperian ocean was extremely 
briny\cite{Kreslavsky:2002,Fairen:2010,Rodriguez:2016}. First, salts would have depressed 
the ocean's water 
triple point by tens of Kelvins, relaxing the constraint on the hypothetical greenhouse gases (assumed here to be CH$_4$ and H$_2$) concentration needed 
for the ocean to remain deglaciated. Next, reducing the ocean surface temperature would reduce the evaporation rate of the ocean, thus 
relaxing the constraint on the rate of replenishment needed for the ocean to survive millions of years. For a drop of 21~K in ocean surface temperature, 
the ocean evaporation rate\cite{Turbet:2017icarus} could drop by a factor of 
$e^{-\frac{L_\text{evap}~\text{M}_{\text{H}_2\text{O}}}{\text{R}}(\frac{1}{252~K}-\frac{1}{273~K})}$~$\sim$~5, 
with $L_\text{evap}$ the latent heat of evaporation, $\text{M}_{\text{H}_2\text{O}}$ the molar mass of water and R the ideal gas constant.
Ultimately, precipitation (here snowfall) near the ocean shorelines would be too cold to melt, thus relaxing the constraint on the 
fluvial erosion. However, for this scenario to work, the salinity of the ocean would have to be extremely high, and 
as much as $\sim$~2$\times$10$^6$~km$^3$ of salts, i.e. $\sim$~10~m GEL (Global Equivalent Layer) would have to be 
accounted for, whatever the nature of brine considered\cite{Mohlmann:2011}. Although salts such as perchlorates have been 
detected in situ by the Phoenix\cite{Hecht:2009} and Viking landers\cite{Navarro-Gonzalez:2010} at the $\sim$~0.1$\%$~level, 
whether this is sufficient or not to account for the presence of $\sim$~10~m GEL of salts is left for future investigations.

\section{Alternative solutions}

Several hypothetical scenarios could potentially reconcile the tsunami hypothesis\cite{Rodriguez:2016,Costard:2017} with 
the geological records and our understanding of the Martian climate.

In one class of scenarios, the ocean existed but it was fully (or almost fully) glaciated, and potentially protected by a lag deposit. Tsunami events could have 
been produced in response to consecutive meteoritic impacts, for instance 
 resulting from a collision with the different pieces of a broken body like 
the Shoemaker-Levy 9 comet which hitted Jupiter in July 1994\cite{Asphaugh:1996}.
The first impact would (at least partially, and as a result of either (i) direct melting or (ii) 
impact-induced global climate change) deglaciate 
the ocean, and the following impacts could then produce the tsunami. 

Tsunami events could also be produced in response to impact-triggered pressure waves propagating in a thin, deep 
subsurface ocean located below a km-thick cover of ice\cite{Rodriguez:2019}. The same impact could also 
expel liquid water from the deep ocean, that would then form a huge flow on top of the ice cover.
In principle, this scenario is compatible with the minimum ice cover thickness of $\sim$~1~km (calculated above for the maximum 
annual mean temperature predicted in the northern lowlands by 3-D Global Climate models\cite{Wordsworth:2013,Turbet:2017icarus}) 
and the maximum depth of the hypothetical ocean\cite{Costard:2017} estimated at $\sim$1.4~km.

In a second class of scenarios, there is no perennial ocean. Instead, the tsunami events could have been produced by 
the catastrophic outflow channel formation events that occurred in the same region and at the same epoch. 
The sudden release of extremely large amounts of water could 
produce large waves across the northern lowlands terrains, and potentially resurfacing them.
A similar scenario was previously invoked to explain the formation of sedimentary deposits on the 
slopes of the northern lowlands\cite{Tanaka:1997}.
For very large discharge rates ($\sim$~10$^9$ m$^3$~s$^{-1}$), 
the mean flow velocity and depth\cite{Turbet:2017icarus} can reach 10~m~s$^{-1}$ and 50~m, respectively, for a flow width 
of $\sim$~2000~km, typical of the Northern lowlands characteristic horizontal extension. 
These numbers are of the same order of magnitude than those calculated for impact-generated tsunami events\cite{Rodriguez:2016,Costard:2017}. 
Although most recorded tsunami deposits do not face the circum-Chryse outflow channels\cite{Rodriguez:2016}, 
these extreme water flows could have been guided in various directions by remnant ice deposits, 
originating either (i) from the freezing of water from previous outflows, or (ii) 
from atmospheric precipitation\cite{Turbet:2017icarus}.


The hypothetical alternative solutions mentioned here 
should be explored in greater detail with appropriate numerical models in the future.

\begin{methods}


\subsection{Ice thickness calculation}
\label{method_ice_calc_nat}

The ice thickness can be estimated using the assumption that the transport of energy inside the water ice 
layer is controlled by conduction. The thermal conduction heat flux $F$ can be written as 
follows:

\begin{equation}
F~=~~\frac{A}{h_{\text{ice}}}~\ln{\Big(\frac{T_{\text{bottom}}}{T_{\text{surf}}}\Big)}
\label{conduc_flux}
\end{equation}
with $\lambda_{\text{ice}}$($T$)~=~$\frac{A}{T}$ the thermal conductivity of ice (with $A$~=~651~W~m$^{-1}$)\cite{Petrenko:2002}, $h_\text{ice}$ the thickness of the water ice layer, $T_{\text{surf}}$ the temperature at the top of the water ice layer and $T_{\text{bottom}}$ the temperature at the bottom. At the interface between ice and liquid water, $T_{\text{bottom}}$ is equal to 273,15~K. At equilibrium, the annual mean thermal conduction heat flux $F$ is dominated by the geothermal heat flux $F_{\text{geo}}$. This results in the expression:
\begin{equation}
h_{\text{ice}}~=~\frac{A}{F_{\text{geo}}}~\ln{\Big(\frac{273.15}{T_{\text{surf}}}\Big)}
\label{tsurf_calc}
\end{equation}

Figure~\ref{fig:ice} presents the minimum ice thickness of the ocean calculated as a function of the 
temperature at the top of the sea ice cover and of the geothermal heat flux.

\subsection{Global Climate Model simulations}
\label{method_GCM_nat}

We use the LMD Early Mars 3-D Global CLimate 
Model\cite{Forget:2013,Wordsworth:2013,Wordsworth:2015,Turbet:2017icarus}. The model includes 
both the water and CO$_2$ cycle (condensation and sublimation on the surface and in the 
atmosphere ; formation and transport of clouds ; precipitation and evaporation). It also 
includes a detailed radiative transfer module adapted to a thick CO$_2$-dominated atmosphere 
complemented with H$_2$O and H$_2$. We performed three numerical climate simulations at the obliquities 
0$^{\circ}$, 20$^{\circ}$ and 40$^{\circ}$. Simulations were performed with a spatial 
resolution of 64x48x26 (in longitude~x~latitude~x~altitude), using the present-day Mars MOLA 
topography. An ocean was placed in the northern lowlands of Mars, at elevations lower than 
-3.9~km. This value was chosen to match the best case of the tsunami propagation scenarios of Costard\cite{Costard:2017}. We use 
a two layers slab-ocean model to treat the oceanic region\cite{Codron:2012}. The transport of heat by the 
ocean circulation is not taken into account here. We fixed the total atmospheric pressure to 
1~bar (see \cite{Kite:2018} and references therein), and varied the concentration of H$_2$ (from 5 to 20$\%$) in order to sustain a 
deglaciated ocean (annual mean temperature is around 276~K). 
The concentration of H$_2$ has been chosen to ensure that the 
deglaciated ocean has the lowest possible temperature.
This was achieved with 
concentrations of H$_2$ equal to 7, 6 and 5~$\%$ for the simulations at 0, 20 and 40$^{\circ}$ 
obliquities, respectively.

Figure~\ref{fig:ocean} presents the annual cumulated rainfall, the annual net surface accumulation of 
water, and the position of permanent ice reservoirs for various 3-D Global Climate simulations.

\end{methods}



\bibliographystyle{naturemag}


\begin{addendum}
 \item  This work was granted access to the HPC resources of the institute for computing and data sciences (ISCD) at Sorbonne Universite. 
This work benefited from the IPSL ciclad-ng facility. We are grateful for the computing resources on OCCIGEN (CINES, French National HPC).
The authors acknowledge Michael Wolff for his useful feedbacks and careful proofreading. 
M.T. acknowledges Jim Head for discussions related to this work, 
as well as Edwin Kite for his useful feedbacks on the manuscript. Eventually, 
the authors acknowledge the reviewer Alexis Rodriguez for his very constructive feedbacks.
 \item[Author Contributions] M.T. developed the core ideas of the manuscript, performed the calculations and the 3-D numerical climate simulations, 
wrote most of the text and prepared Figures~\ref{fig:ice}-\ref{fig:ocean}. F.F. contributed to the design and structure of the manuscript, and reviewed the manuscript.
 \item[Competing Interests] The authors declare that they have no competing interests. 
 \item[Correspondence] Correspondence and requests for materials
should be addressed to M.T.~(email: martin.turbet@lmd.jussieu.fr).
\end{addendum}

\end{document}